\documentclass[12pt,preprint]{aastex}

\usepackage{graphicx,epsfig,subfigure}
\usepackage{amssymb,amsmath}
\usepackage{epstopdf}

\slugcomment{To appear in ApJL}

\shorttitle{Detection of FRBs with the MWA}
\shortauthors{Trott, Tingay \& Wayth}

\begin{document}


\title{Prospects for the Detection of Fast Radio Bursts with the Murchison Widefield Array}


\author{Cathryn M. Trott\altaffilmark{1,2}, Steven J. Tingay\altaffilmark{1,2} and Randall B. Wayth\altaffilmark{1}}
\affil{International Centre for Radio Astronomy Research, Curtin University, Bentley WA 6845, Australia}
\email{cathryn.trott@curtin.edu.au}


\altaffiltext{2}{ARC Centre of Excellence for All-Sky Astrophysics (CAASTRO)}


\begin{abstract}
Fast Radio Bursts (FRBs) are short timescale ($<<$1~s) astrophysical radio signals, presumed to be a signature of cataclysmic events of extragalactic origin. The discovery of six high-redshift events at $\sim$1400~MHz from the Parkes radio telescope suggests that FRBs may occur at a high rate across the sky. The Murchison Widefield Array (MWA) operates at low radio frequencies (80-300~MHz) and is expected to detect FRBs due to its large collecting area ($\sim$2500~m$^2$) and wide field-of-view (FOV, $\sim$1000~square degrees at $\nu$=200~MHz). We compute the expected number of FRB detections for the MWA assuming a source population consistent with the reported detections. Our formalism properly accounts for the frequency-dependence of the antenna primary beam, the MWA system temperature, and unknown spectral index of the source population, for three modes of FRB detection: coherent; incoherent; and fast imaging. We find that the MWA's sensitivity and large FOV combine to provide the expectation of multiple detectable events per week in all modes, potentially making it an excellent high time resolution science instrument. Deviations of the expected number of detections from actual results will provide a strong constraint on the assumptions made for the underlying source population and intervening plasma distribution.
\end{abstract}


\keywords{instrumentation: detectors --- methods: data analysis --- methods: observational --- radio continuum: general
 --- surveys}

\section{Introduction}

The first evidence for a new class of short duration radio transients emerged in 2007, with the detection of a highly dispersed single pulse of radio emission with the Parkes radio telescope by \citet{lorimer07}, who suggested that the burst was caused by an unknown form of explosive event at a cosmological distance.  This interpretation could not be verified, because the poor angular resolution of Parkes did not admit the possibility of localizing the emission well enough to identify the progenitor or its (presumed) host galaxy.

In the years since, the evidence for a new class of such bursts and the interpretation advanced by \citet{lorimer07} has waxed and waned \citep{keane10, keane11, burkespolaor11, bannister12}.  Most recently, results that strongly suggest a new population of explosive events at cosmological distances exists have been presented \citep{thornton13}.  While physical interpretations for these events remain speculative, they are thought to involve highly compact astrophysical objects \citep[][and references therein]{falcke13}.  \citet{thornton13} coin the term Fast Radio Bursts (FRBs) to describe the events detected thus far.

All known FRBs have been detected thus far by the Parkes radio telescope, however.  To further verify an astronomical origin for FRBs, comfort would be found in detecting FRBs using a diversity of radio telescopes.  Furthermore, the limited angular resolution of Parkes means that none of the FRBs detected thus far have been localized or associated with progenitor objects or host galaxies. The next step toward understanding FRBs lies in detecting more events using a range of instruments and, most importantly, localizing the events for identification purposes.  A range of experiments have been performed or are underway to meet this challenge \citep[see][and references therein]{trott13_vfastr,wayth12}.

In addition to the experiments and results presented to date, there is a growing awareness of the impact the new generation of low frequency radio telescopes may be able to make in further understanding FRBs. \citet{lorimer13} remark that observed FRBs appear to fall below the Galactic DM-SM relation. \citet{macquart13} propose that the scattering of the radio emission by the intergalactic medium may be less of an effect than previously thought, making low frequency detection experiments viable.  Efforts are being put into observational programs at frequencies below 1~GHz, for example using the Square Kilometre Array Molonglo Prototype (SKAMP: Bailes, M., private communication) and LOFAR \citep{falcke13b}.

Given the suggestion of \citet{lorimer13}, we examine here the potential for another new low frequency radio telescope, the Murchison Widefield Array (MWA) \citep{tingay13_mwasystem}, to detect and localize FRBs. We extend the work of \citet{hassall13}, who predict FRB detection rates for a range of instruments using broad parameters. The advantages of the MWA for FRB searches are: extremely large fields of view; reasonable angular resolution for localization and identification ($\sim$arcminute); flexible observation modes that allow incoherent and coherent searches; large amounts of available observing time; and a site that is virtually free of human-made radio interference (RFI).  In this Letter, we predict the detection rates of FRBs with the MWA under a range of assumed conditions and for a range of possible observational modes.

We introduce the MWA and its FRB detection modes, before describing the detection framework developed in \citet{trott13_vfastr}, which computes the signal-to-noise ratio (S/N) of a signal at any location in the MWA primary beam. This formalism allows us to compute the total sky area above a given S/N threshold for a variety of observational and source parameters. We then assume a simple model for the FRB source population, based on a population of standard candles, as described by \citet{lorimer13}, yielding the sky density of sources as a function of redshift and flux density. The product of the MWA sky area and source density yields the expected number of sources as a function of redshift.

\section{Methods}
\subsection{Cold plasma dispersion}
Interaction of the wavefront with electrons along the line-of-sight (LOS) between the source and observer, produces a frequency-dependent delay in the arrival of the wavefront:
\begin{equation}
\Delta{t}(\nu) = \frac{e^2}{2\pi{m_e}c}\frac{1}{\nu^2}\int n_e {\rm d}l = 4.15 \frac{{\rm DM}}{\nu^2} {\rm ms},
\label{cold_plasma_dispersion}
\end{equation}
where the integral of the electron density, DM, is the dispersion measure (pc.cm$^{-3}$). This dispersive delay is fundamental for FRB searches, allowing discrimination between astrophysical sources and local transient phenomena (e.g., RFI).

\subsection{The Murchison Widefield Array}
The Murchison Widefield Array \citep[MWA,][]{tingay13_mwasystem} is a 128-tile low-frequency radio interferometer located in Western Australia.  The MWA architecture includes the Voltage Capture System (VCS), a sub-system that allows the Nyquist-sampled and digitized voltage representation of the waveforms incident on the 128 tiles, for the full bandwidth (over 3072$\times$10 kHz channels) and for both linear polarizations, to be recorded to computer hard disk.  The VCS allows the rawest form of data to be collected from the MWA system for offline processing, and is described in \citet{tingay13_mwasystem}.

Detection of FRBs with the MWA is a three-step process, and three modes of detection are employed: coherent; incoherent; and fast imaging.  The voltage outputs of the VCS are combined across tiles and powers are formed from the combined voltages, depending on which of the three detection modes is being used:
\begin{itemize}
\item Coherent -- Voltages are summed across tiles, after application of geometric and calibration delays to point on the sky. The coherent power is formed by squaring after the summation of signals. This mode carries phase information from each tile. It yields the smallest FOV per beam (the tied-array beam, equivalent to the synthesized beam, $\propto\lambda/D$, where $D$ is the length of the longest baseline in the array, $\sim$3~km, and $\lambda$ is the wavelength of the radiation), and the largest sensitivity ($\propto N_{\rm ant}$). Here we regard the coherent mode as tiling the full primary beam with tied-array beams, thereby providing the upper limit to detectability with the MWA;
\item Incoherent -- Voltages are squared independently for each tile, and then summed to form the total incoherent power. This mode carries no phase information from each tile. It is the simplest to implement, yields the largest FOV (the individual tile beam, $\propto\lambda/d$, where $d$ is the size of the tile, $\sim$5~m), but has the lowest sensitivity ($\propto\sqrt{N_{\rm ant}}$);
\item Imaging -- Voltages are cross-correlated between tiles in the normal interferometric mode, but with fine time resolution ($\sim$50~ms), followed by Fourier transform of each spectral channel to the image-plane. This mode yields moderate sensitivity ($\propto\sqrt{N_{\rm ant}(N_{\rm ant}-1)}$) and a large FOV (the tile beam) with good angular resolution (the synthesized beam).
\end{itemize}

Dynamic spectra are then formed (power in each spectral and temporal channel). For the imaging mode, dynamic spectra are formed for each pixel in the image plane.  Data are de-dispersed incoherently by temporal shifting of power samples and summation across frequency, and a threshold is applied to detect any signals above the noise level.
Table \ref{detection_parameters} lists the parameters for each observing mode. The final column describes how the algorithmic load (operations per second) scales with observing parameters.
\begin{table}[ht]
\begin{minipage}{15cm}
\begin{tiny}
\begin{tabular}{cccccc}
\hline Mode & $f(N)$ & $\Delta{t}_{s}$ & $\Delta{\nu}$ & BW & Cost\footnote{$N_{\rm p}=(D/d)^2$ (number of pixels, or tied-array beams); $N_{\rm BL}=N_{\rm ant}(N_{\rm ant}-1)/2$ (number of baselines); $N_{\rm grid}\sim{10}$ (gridding kernel size); R (rate: images per second); $N_{\rm DM}\sim{1000}$ (number of DM trials); $N_{\rm pol}={2}$ (number of polarizations); and $N_{\rm chan}$ (number of spectral channels). Values assume full use of instrument; for the MWA, removal of long baselines can potentially reduce the number of image-plane pixels, with little loss of sensitivity, reducing the cost.} \\ 
Unit & & (ms) & (kHz) & (MHz) & (ops $s^{-1}$) \\
\hline Coherent & 128 & 0.1 & 10 & 30.72 & ${\rm BW}N_{\rm ant}N_{\rm pol}N_{\rm p}N_{\rm DM}\sim 10^{18}$ \\
Incoherent &	$\sqrt{128}\simeq{11.3}$ & 0.1 & 10 & 30.72 & ${\rm BW}N_{\rm ant}N_{\rm pol}N_{\rm DM}\sim 10^{13}$ \\
Imaging &	$\sqrt{128.127}\simeq{128}$ & 50 & 40 & 30.72 & ${\rm R}N_{\rm chan}(N_{\rm BL}N_{\rm grid}N_{\rm pol}+N_{\rm p}\log_2{N_{\rm p}}+N_{\rm p}N_{\rm DM})\sim 10^{13}$ \\
\hline 
\end{tabular}
\caption{Observational parameters used for the three FRB modes with the MWA, and approximate algorithmic load assuming use of the full array.}\label{detection_parameters}
\end{tiny}
\end{minipage}
\end{table}


\subsection{Detection metric}
Detection of FRBs from dynamic spectra is comprised of: (1) de-dispersion of the power samples in the dynamic spectrum according to Equation \ref{cold_plasma_dispersion}; (2) summation over spectral channels to form a power time series; (3) estimation of the sample noise level (root-mean-square); and (4) thresholding of the data at a specified S/N. For detection of FRBs, the DM is unknown \textit{a priori}, and many trial DM values are applied to the data. In \citet{trott13_vfastr}, we extended the common formalism to develop a detection metric that correctly incorporates frequency-dependent beam shapes, frequency-dependent noise levels (System Equivalent Flux Density, SEFD) and the spectral shape of the target signals (spectral index). This formalism yields significant departures from the assumptions of a fixed beam and noise properties for large fractional bandwidth experiments and wide-field instruments. We refer the reader to this work for further details.

We write the expected S/N for an event, as a function of location on the sky with respect to the primary pointing centre ($\theta,\phi$), and parametrized by the source flux density ($S_0(\nu_0)$), spectral index ($\alpha$), system bandwidth (BW), spectral resolution ($\Delta\nu$), and temporal sampling timescale ($\Delta{t}_s$). This expression is:
\begin{equation}
{\rm S/N}(\theta,\phi;\alpha,{\rm BW},\Delta\nu,\Delta{t}) = f(N) g(\tau) S_0(\nu_0) \frac{\displaystyle\int_{\rm BW} \left(\frac{\nu}{\nu_0}\right)^\alpha B(\nu;\theta,\phi){\rm d}\nu} {\sqrt{ \Delta\nu \displaystyle\int_{\rm BW} \sigma^2(\nu;\Delta\nu,\Delta{t}_s){\rm d}\nu }},
\label{snr_equation}
\end{equation}
where,
\begin{equation}
S(\nu) = S_0(\nu_0) \left(\frac{\nu}{\nu_0}\right)^\alpha \,\,\,{\rm Jy},
\end{equation}
is the source flux density at frequency, $\nu$, normalized to $S_0$ (Jy) at the reference frequency, $\nu_0$. The frequency-dependent primary beam shape, $B(\nu)$ is normalized to unity at zenith. The system noise, $\sigma(\nu;\Delta\nu,\Delta{t}_s)$ is the flux density uncertainty in each spectral and temporal channel, and is given by:
\begin{equation}
\sigma(\nu;\Delta\nu,\Delta{t}_s) = \frac{\text{SEFD}}{\sqrt{2\Delta\nu \Delta{t}_s}} = \frac{2kT_{\rm sys}}{A_e} \frac{1}{\sqrt{2\Delta\nu \Delta{t}_s}} \,\,\, {\rm Jy},
\end{equation}
where $A_e$ is the effective antenna area, $T_{\rm sys}$ is the system noise, and two instrumental polarizations have been considered. At low frequencies, the system noise temperature is dominated by sky emission. Off the plane of the Galaxy, the sky temperature can be approximated with \citep{fob06}:
\begin{equation}
T_{\rm sys} \simeq 180 \left( \frac{\nu}{180{\rm MHz}} \right)^{-2.6} \,\,\, {\rm K}.
\end{equation}
The functions $f(N)$ and $g(\tau)$ are the dependence of the system sensitivity on the number of antennas and temporal sampling, respectively, and differ for the observation modes:
\begin{equation}
  f(N)=\begin{cases}
    N_{\rm ant}, & \text{Coherent}.\\
    \sqrt{N_{\rm ant}}, & \text{Incoherent} \\
    \sqrt{N_{\rm ant}(N_{\rm ant}-1)}, & \text{Imaging}.
  \end{cases}
\end{equation}
The pulse energy is contained within an intrinsic timescale, $W_{\rm int}$, and the signal is degraded linearly if the sampling timescale exceeds this scale. The signal is further degraded by the finite spectral resolution of the instrumental signal processing, and by scattering of the wavefront due to turbulence in the intervening plasma. Scattering of the signal temporally broadens the pulse, but the degree to which this scattering occurs is uncertain. \citet{cordes03} provide an empirical relation for the scattering due to the Galaxy, but \citet{thornton13} find mixed evidence for scattering in their sample of extragalactic sources. As such, we take two extremes in this work: `zero scattering' and `high scattering', with the latter given by the empirical Galactic expression.
\vspace{0.2cm}
\newline\noindent
{\bf Zero scattering: }
Due to the finite spectral resolution of the dynamic spectrum (channelization of the data) and the dispersive delay across a single channel, the signal observed by the system has a characteristic scale,
\begin{equation}
W_{\rm obs} = \sqrt{W_{\rm int}^2 + \Delta{t}_{\rm DM}^2},
\end{equation}
where,
\begin{equation}
\Delta{t}_{\rm DM} \simeq 8.3 \frac{{\rm DM}}{\nu_{\rm GHz}^3}\Delta{\nu}_{\rm MHz} \,\,\, {\rm \mu{s}}.
\label{temporal_smearing}
\end{equation}
We assume that signals that have been smeared across multiple temporal samples are recovered by averaging over adjacent temporal bins (boxcar averaging), yielding an overall reduction in sensitivity proportional to the square-root of the averaging timescale. The sensitivity function is therefore,
\begin{equation}
  g(\tau)=\begin{cases}
    \frac{W_{\rm int}}{\Delta{t}_s}, & \text{if } W_{\rm obs}<\Delta{t}_s\\
    \sqrt{\frac{W_{\rm int}}{\Delta{t}_s}}, &  \text{if } W_{\rm int}>\Delta{t}_s\\
    \frac{W_{\rm int}}{\sqrt{W_{\rm obs}\Delta{t}_s}}, & \text{if } W_{\rm obs}>\Delta{t}_s \text{ and } W_{\rm int}<\Delta{t}_s.
  \end{cases}
\end{equation}
\newline\noindent
{\bf High scattering: }
\citet{cordes03} and \citet{bhat04} describe the Galactic scattering timescale, $\tau$ ($\mu$s), parametrically;
\begin{equation}
\log{\tau} = -3.72 + 0.411 \log{\rm DM} + 0.937(\log{\rm DM})^2 - 3.9\log{\nu_{\rm GHz}}.
\label{tau_d}
\end{equation}
The S/N degradation due to convolution of a square pulse with an exponential scattering tail with characteristic timescale $\tau$ is given by \citet{trott13_vfastr}:
\begin{equation}
\epsilon \equiv \frac{SNR_{\tau}}{SNR_{\rm optimal}} = \sqrt{1-\beta+\beta\exp{(-1/\beta)}},
\end{equation}
where $\beta = \tau/W_{\rm obs}$. The sensitivity function, $g(\tau)$, for the high scattering case is degraded by the parameter $\epsilon$ compared with the zero scattering case.

Note that although we present these two extremes in this work, the evidence of \citet{thornton13} and discussion by \citet{lorimer13} and \citet{macquart13} suggest that reality is closer to the zero scattering case.

\subsection{Source population}
Predictions for the source population of FRBs are difficult to make due to the small number of detections to date, and the inability to localize their host galaxies. This leads to uncertainties in the luminosity distribution and the plasma properties along the LOS. \citet{lorimer13} modeled the known FRB population as standard candles, and computed the intrinsic bolometric luminosity of FRBs in the frequency range 10~MHz$-$10~GHz, under this assumption:
\begin{equation}
L_{\rm int} = 8 \times 10^{37} \,\,\, {\rm J.s^{-1}},
\end{equation}
from which the peak flux density, $S_0$, can be derived for a given redshift $z$. We use their calibrated curve to predict the peak flux density at 1400~MHz and extrapolate to lower frequencies using an observed spectral index, $\alpha$, noting that their calibration relies on an assumed spectral index of $-1.4$. There is a strong dependency of the predicted detection rates on the unknown spectral index of the source, and we compute expected rates for three values: $\alpha$ = $-2$; $-1$; and 0.

\citet{lorimer13} also provide an empirical fit to the cumulative rate density of events as a function of redshift:
\begin{equation}
R(<z) \simeq \left( \frac{z^2 + z^3}{4} \right) \,\,\, {\rm day^{-1} deg^{-2}}.
\end{equation}
Together these provide the number density of events per unit time and area, and the pulse energy, both as a function of redshift. Finally, to compute the dispersion measure, DM, as a function of redshift, we follow \citet{inoue04} and assume that the IGM is filled uniformly with ionized plasma at the measured WMAP mean density \citep{bennett12}:
\begin{equation}
{\rm DM}_{\rm IGM}(z) = \displaystyle\int_0^z {\rm d}z^\prime \frac{c}{H_0}\frac{n_{e,0}(1+z^\prime)}{\sqrt{\Omega_m(1+z^\prime)^3+\Omega_\Lambda}},
\end{equation}
where $n_{e,0}=2.1\times{10^{-7}}$~cm$^{-3}$ is the mean baryon density, $\Omega_m=0.27$ is the matter energy density, and $\Omega_\Lambda=0.73$ is the dark energy density. In addition to the redshift-dependent component of the DM value, there are contributions from the Galaxy (fixed) and host galaxy. For the Milky Way (MW), we are considering LOS away from the Galactic plane, and take an average of the DM contribution over the sky area above a Galactic latitude $b=5$~degrees, using a parametric fit to the NE2001 model \citep{cordes_ne2001_03}, yielding 64~pc cm$^{-3}$. For the host galaxy contribution, we consider an identical galaxy to the MW (with an average contribution over all LOS), and place it at cosmological distance, yielding an overall DM budget:
\begin{equation}
{\rm DM}_{\rm Tot}(z) = {\rm DM}_{\rm MW} + {\rm DM}_{\rm gal}(z) + {\rm DM}_{\rm IGM}(z) = 64 + \frac{100}{1+z} + {\rm DM}_{\rm IGM}(z).
\end{equation}

\section{Results}
Table \ref{results_table} lists the expected number of detections per day for the MWA (assuming 10 hours of zenith-pointed observation per calendar day), summed over redshift, and including stochastic uncertainties (described below).
\begin{table}
\begin{center}
\begin{tabular}{ccccc}
\hline
$\alpha$ & Scatter & Coherent & Incoherent & Imaging\\\hline 
$-2$ & Zero & 88$\pm$19 & 16$\pm$8 & 38$\pm$12\\
$-1$ & Zero & 23$\pm$9 & 3.5$^{+3.0}_{-3.5}$ & 8.5$^{+5.0}_{-6.0}$\\
$0$ & Zero & 5.6$^{+4.4}_{-5.6}$ & 0.7$^{+1.5}_{-0.7}$ & 1.7$^{+1.8}_{-1.7}$\\
$-2$ & High & 8.3$^{+4.9}_{-5.9}$ & 1.7$^{+1.8}_{-1.7}$ & 3.3$^{+3.0}_{-3.3}$\\
$-1$ & High & 2.5$^{+3.0}_{-2.5}$ & 0.4$^{+1.0}_{-0.4}$ & 0.8$^{+1.5}_{-0.8}$\\
$0$ & High & 0.6$^{+1.4}_{-0.6}$ & 0.1$^{+0.2}_{-0.1}$ & 0.2$^{+0.5}_{-0.2}$\\\hline
N$_{\rm noise}$ ($>7\sigma$) && $2\times{10}^5$ & 0.5 & 300\\
N$_{\rm noise}$ ($>8\sigma$) && 80 & 2$\times$10$^{-4}$ & 0.2\\
\hline
\end{tabular}
\caption{Expected number of fast transient detections per 10-hour day with S/N~$\geq{7}$ for each observing mode of the MWA, for zero-scatter and high-scatter scenarios, assuming ten hours per night of zenith observing. Uncertainties describe the 68\% confidence intervals for a single night of observing. Also listed are the expected number of detections due to noise, N$_{\rm noise}$. For the coherent case, a higher threshold of 8$\sigma$ is more feasible.}\label{results_table}
\end{center} 
\end{table}
As expected, the fully coherent mode, which tiles the \textit{entire} primary beam, yields the highest estimates. The more computationally realistic modes (incoherent and fast imaging) yield lower rates of expected events. However, even with these modes, several events per week would be expected. Assuming a detection threshold of S/N~$=7$ and 1024 independent DM trial values yields feasible expectations of detections per day due to noise for the incoherent and imaging cases. For the fully coherent mode, where there are $\sim{10}^6$ tied-array beams tiling the primary beam, a higher threshold would be more appropriate (for $C=8$, the expected number of detections reduces marginally to 81 from 88).  

\begin{figure}
\begin{center}
\subfigure[Coherent.]{\includegraphics[scale=0.45]{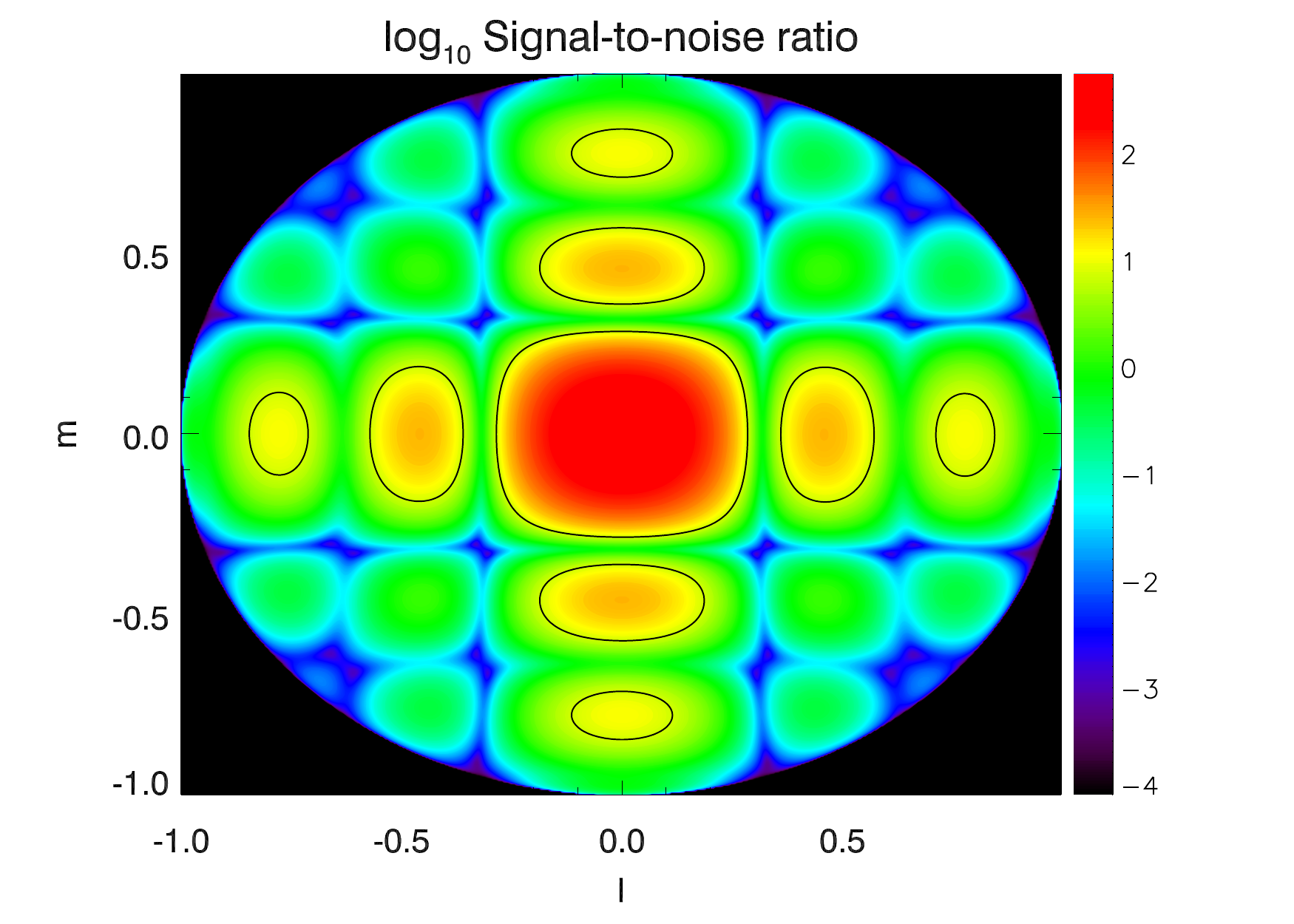}}
\subfigure[Imaging.]{\includegraphics[scale=0.45]{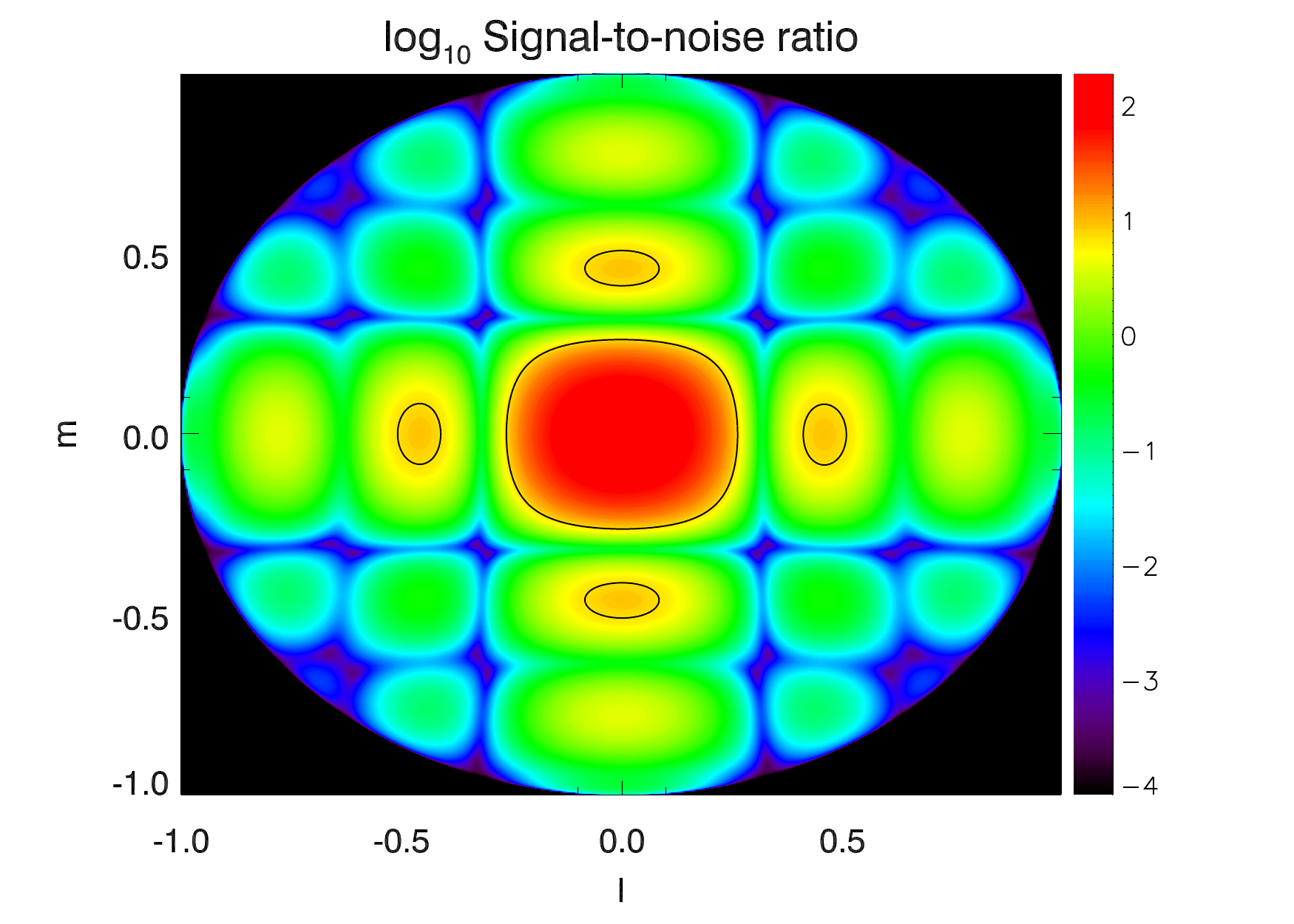}}
\caption{Logarithm of the S/N as a function of sky position (sine projection) for two observing modes at $z=0.2$ and with $\alpha =-2$. The black contour delineates the S/N=7 threshold.}\label{example_beams_fig}
\end{center}
\end{figure}

The wide field-of-view of the MWA, and the non-zero response at large angles from the boresight lead to a large fraction of the sky being available for detection for high S/N events, such as those expected at low redshift. Figure \ref{example_beams_fig} demonstrates this large field-of-view by displaying the S/N as a function of sky position (in sine projection) for an example $z=0.2$ event with spectral index, $\alpha =-2$.
Both the coherent and imaging modes are shown, and the black contour marks the region with S/N~$>$~7. High S/N events are detectable outside of the main lobe of the primary beam.

\begin{figure}
\begin{center}
\subfigure[Coherent.]{\includegraphics[scale=0.45]{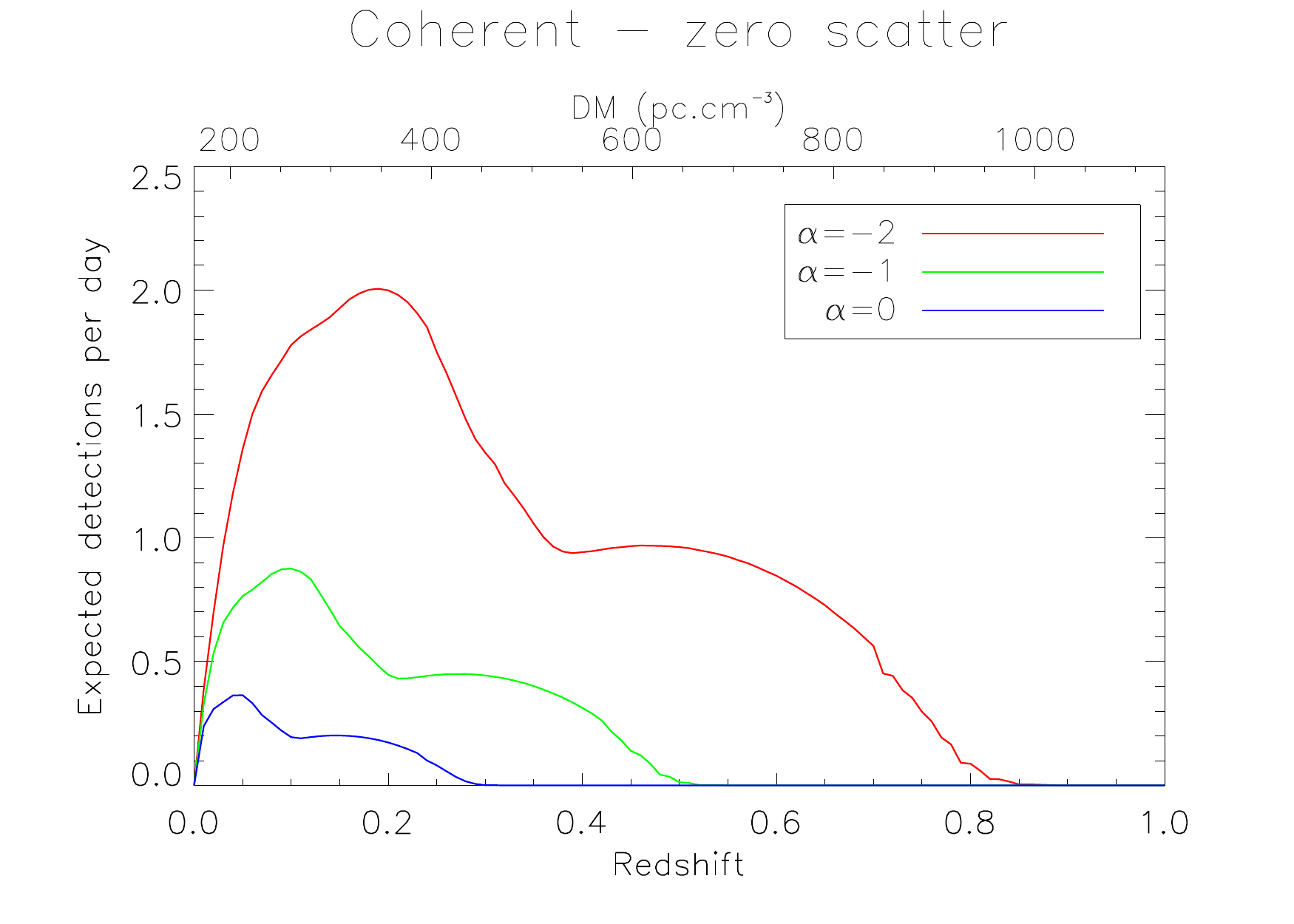}}
\subfigure[Coherent.]{\includegraphics[scale=0.45]{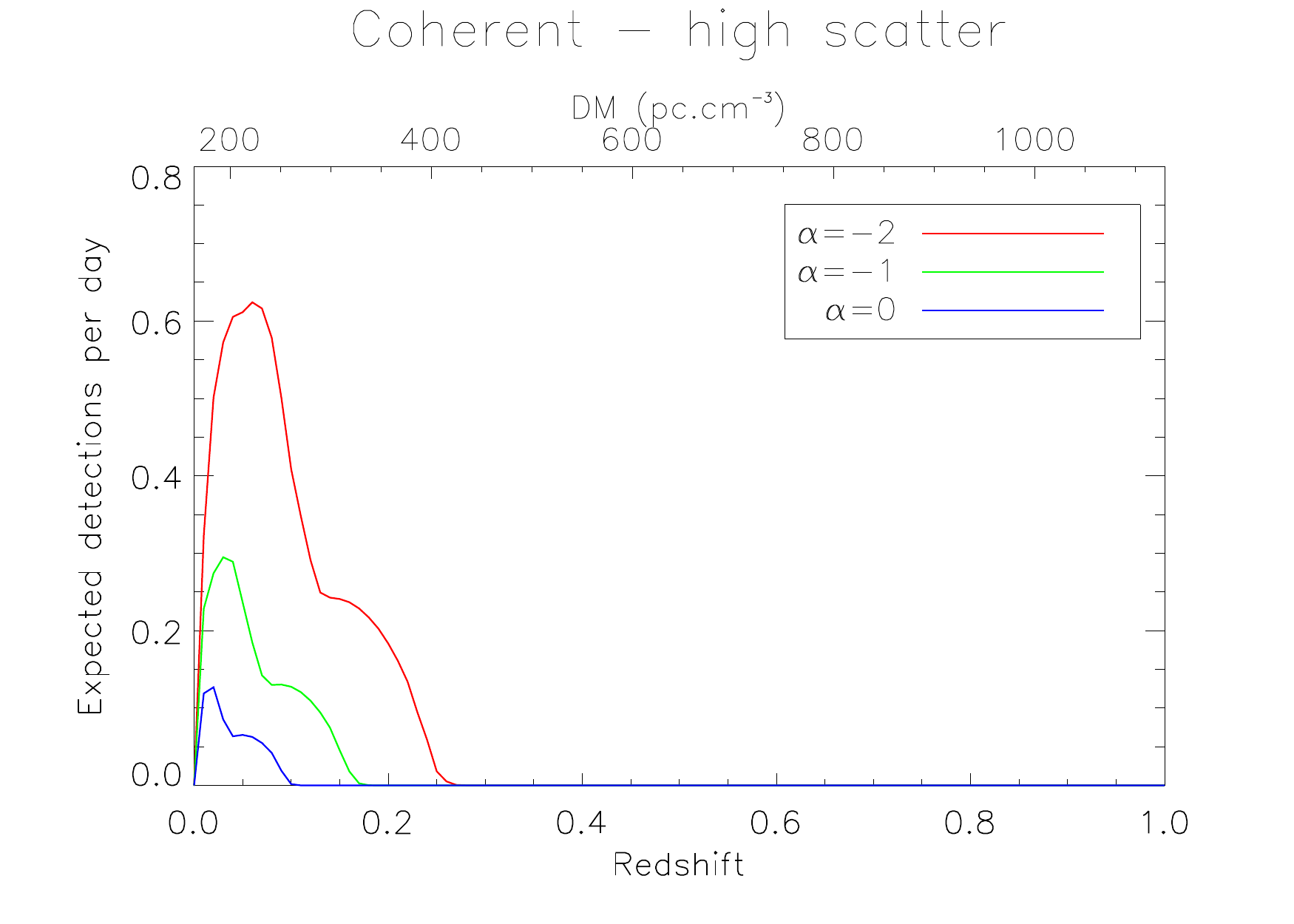}}\\
\subfigure[Incoherent.]{\includegraphics[scale=0.45]{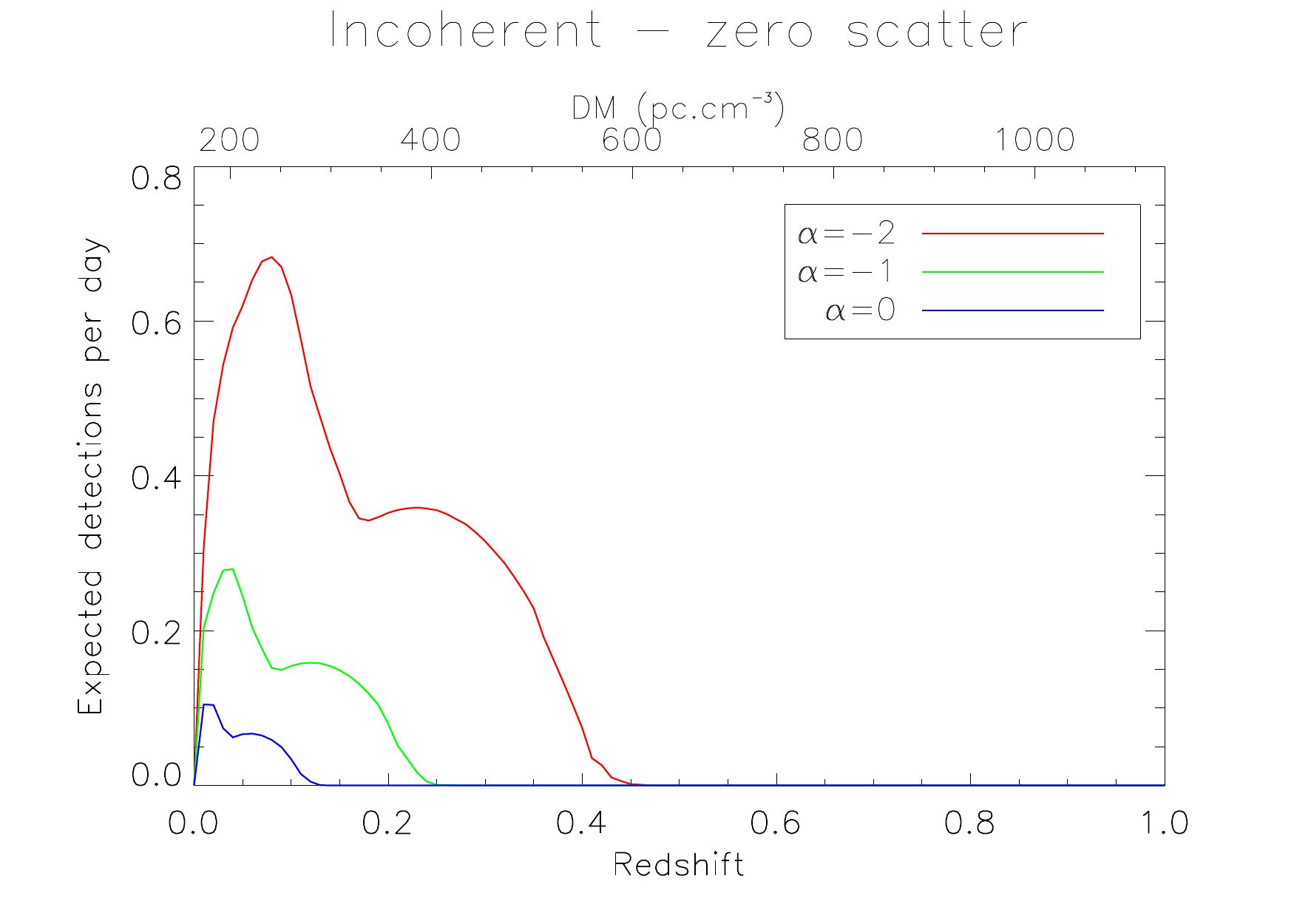}}
\subfigure[Incoherent.]{\includegraphics[scale=0.45]{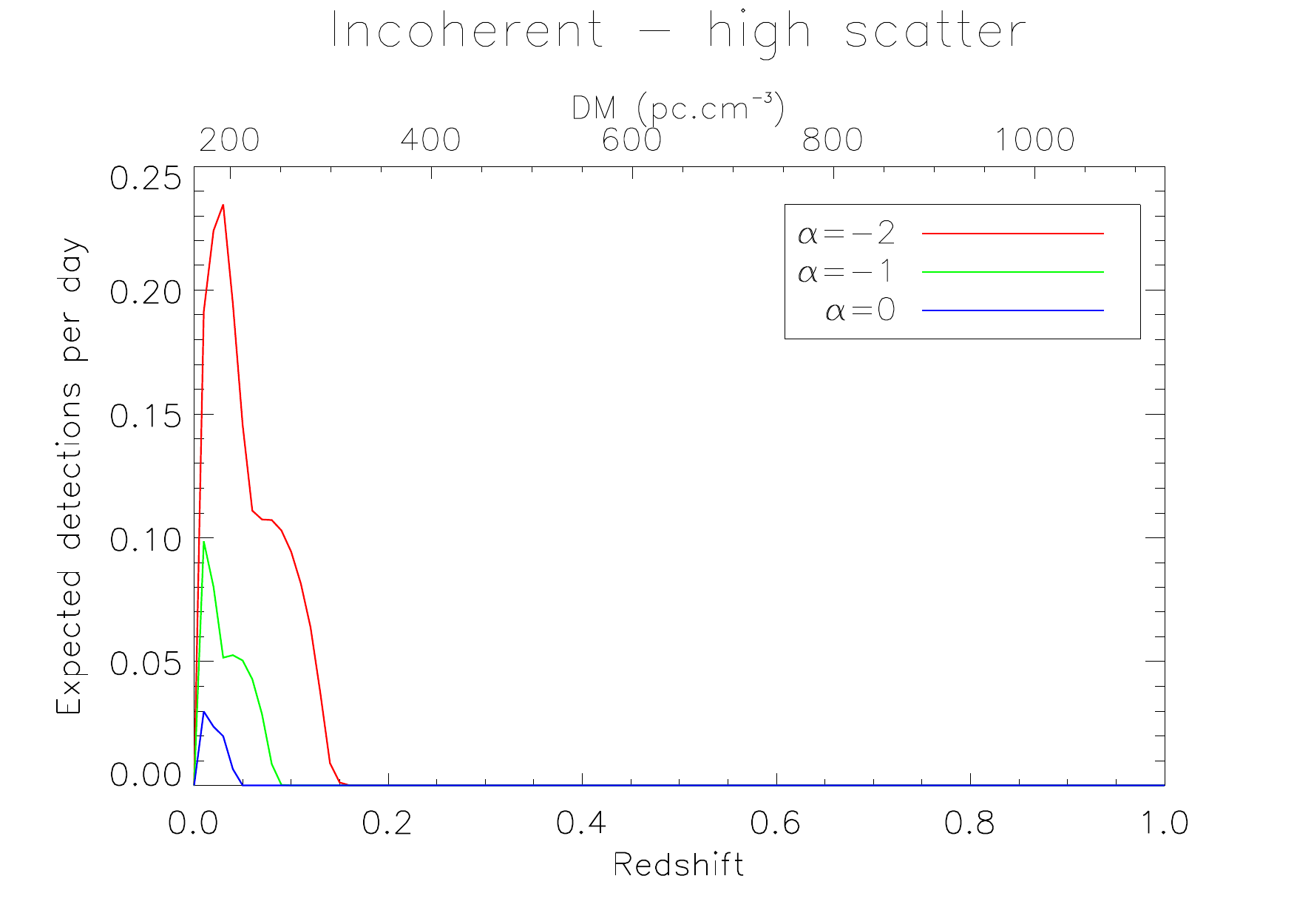}}\\
\subfigure[Fast imaging.]{\includegraphics[scale=0.45]{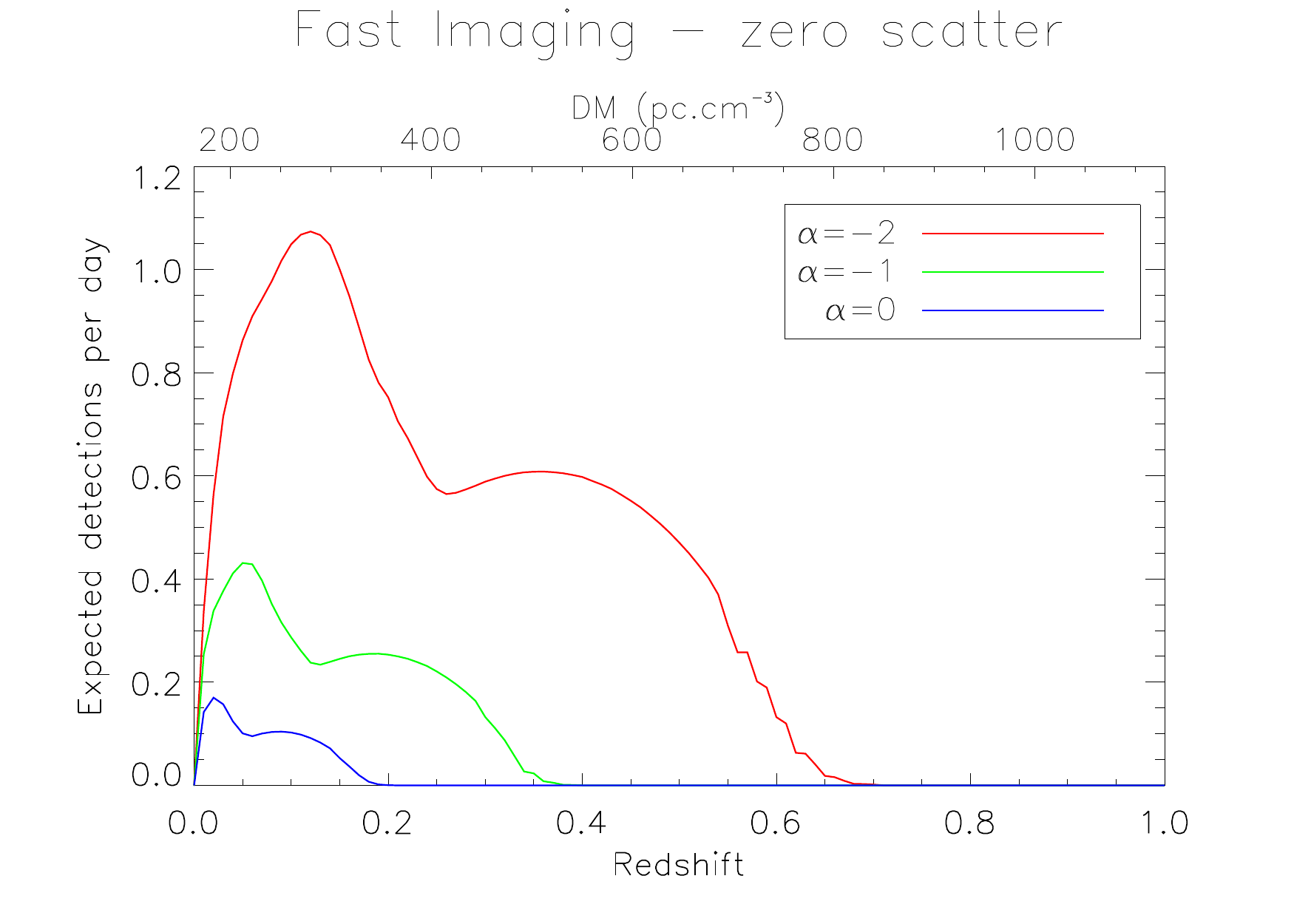}}
\subfigure[Fast imaging.]{\includegraphics[scale=0.45]{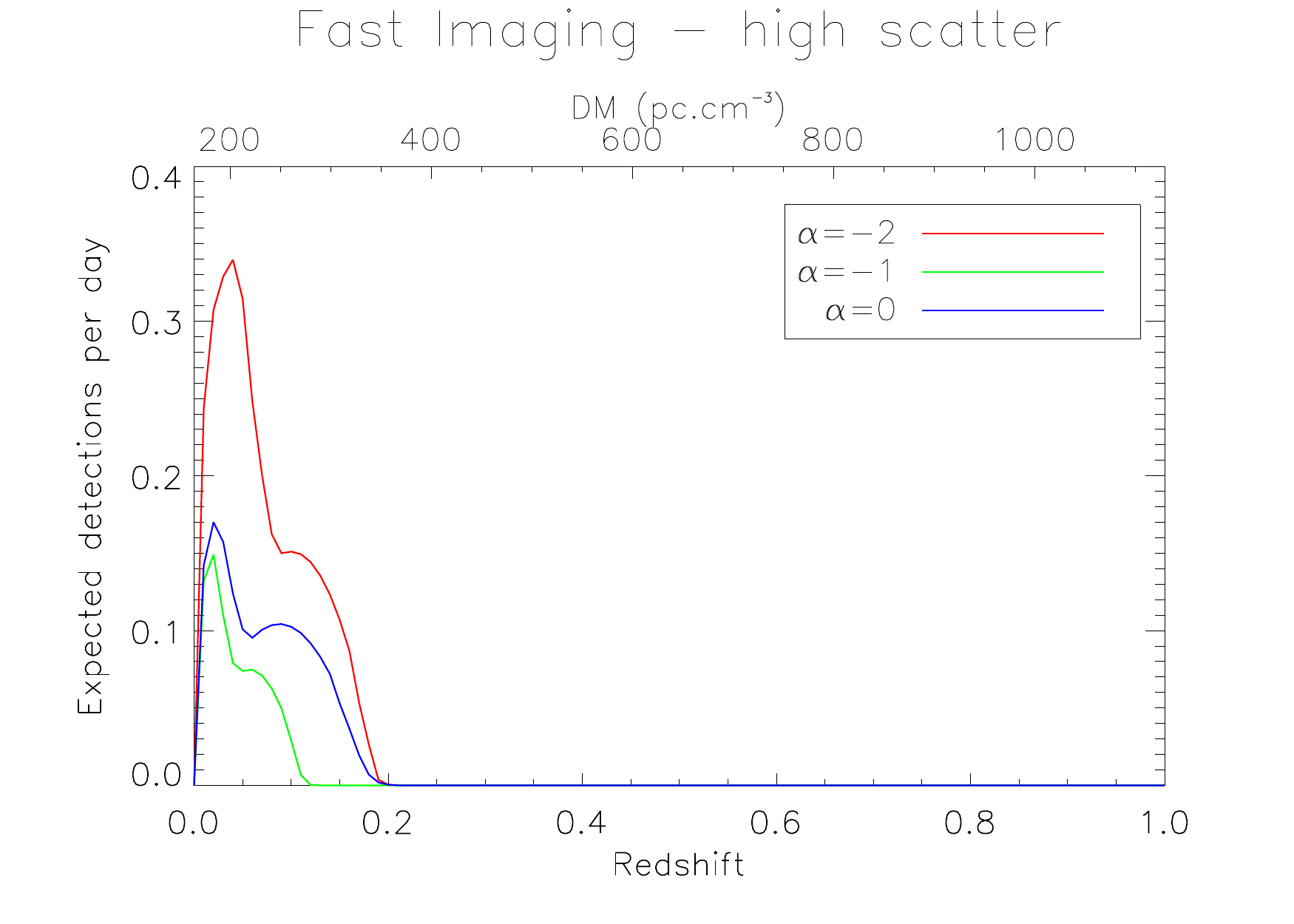}}
\caption{Expected number of detections per day per $\Delta{z}=0.01$ redshift bin, for each observing mode and three values for the spectral index, $\alpha$. Both the zero (left) and high scattering (right) cases are treated.}\label{expected_detections_fig}
\end{center}
\end{figure}

Figure \ref{expected_detections_fig} displays the expected number of events per day per $\Delta{z}=0.01$ redshift bin, as a function of redshift for each mode, spectral index and scattering regime. 
The inflection points in the profiles are due to the gradient in the beam area as a function of threshold (see beam shape in Figure \ref{example_beams_fig}).
The reduced sensitivity of the incoherent mode and temporal resolution degradation of the imaging mode lower the high redshift detection rates, but these may be recovered in the coherent mode for low scattering regimes. The high scattering regime provides a lower limit to the expected number of detections, and yields zero detections with $z>0.3$ under these assumptions and source model. 

\begin{figure}
\begin{center}
\includegraphics[scale=0.7]{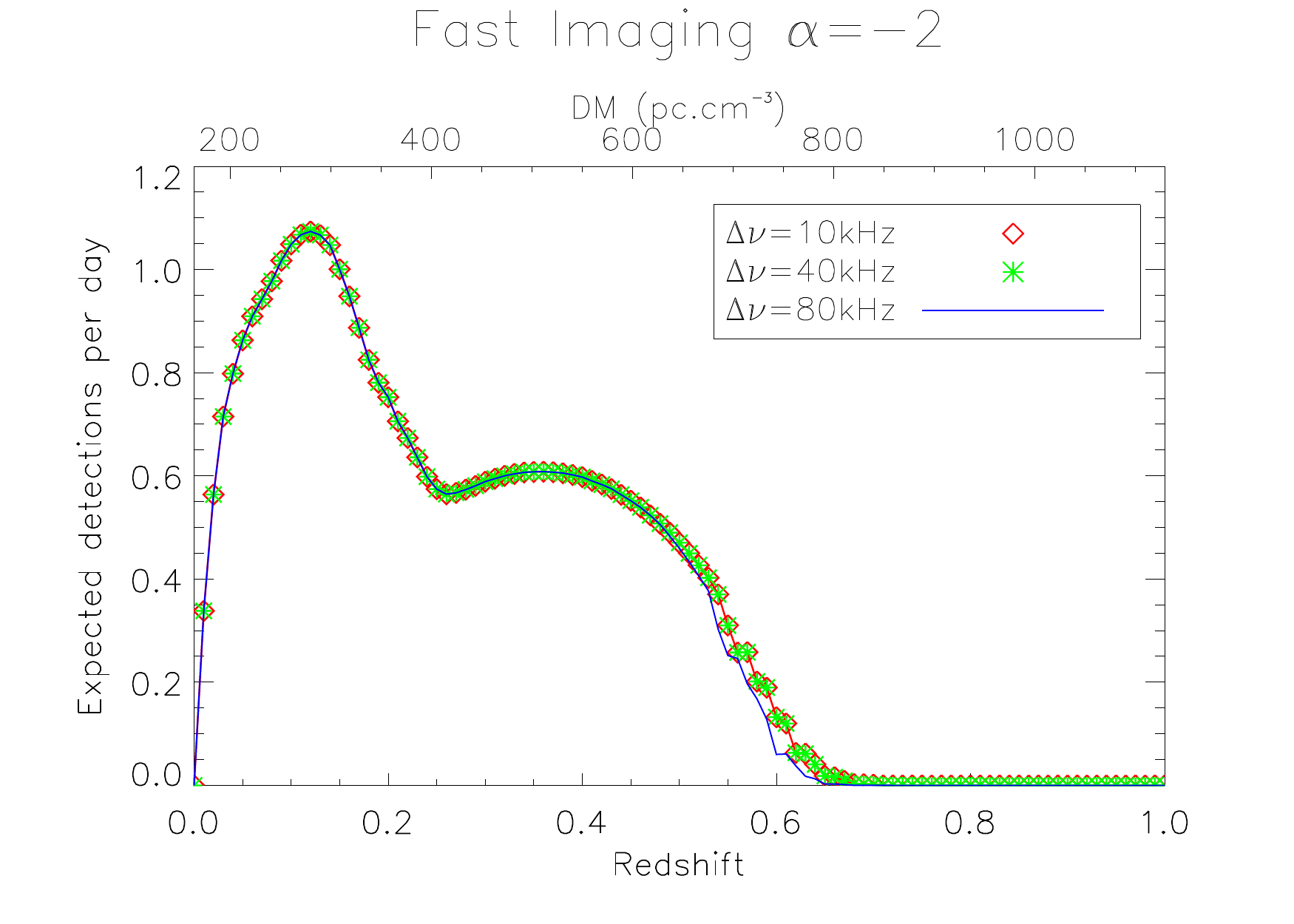}
\caption{Expected number of detections per day for the fast imaging mode and three values for the spectral resolution.}\label{imaging_resolution_fig}
\end{center}
\end{figure}

The fast imaging mode is limited by temporal and spectral resolution, which degrade the S/N by smearing the signal energy over time. Detectability can be improved by increasing the spectral resolution, but this improvement is limited by the native temporal resolution of the system, beyond which temporal smearing due to finite channel width is not the limiting factor. At 200~MHz and 40~kHz spectral resolution, the channel smearing is less than the 50~ms temporal resolution for DM~$\lesssim$~1200\footnote{
$\Delta{t}_{\rm DM} = 41.5\left(\frac{{\rm DM}}{1000}\right)\left(\frac{\Delta\nu}{0.04{\rm MHz}}\right)$ ms
} (equation \ref{temporal_smearing}). Figure \ref{imaging_resolution_fig} demonstrates this effect by comparing detection performance for three values of the spectral resolution.
Channels of width 10~kHz and 40~kHz show indistinguishable results, and performance is only degraded at higher redshifts for 80~kHz channels.

\subsection{Uncertainties}
There are two sources of stochastic uncertainties in the calculations presented here: (1) variability in the rate density of sky events; and (2) uncertainty in the detectability of a source due to noise fluctuations (effectively varying the area of the sky available for a 7$\sigma$ detection). There are substantial systematic errors associated with the source and plasma distribution assumptions, most prominently in the unknown luminosity function, but they will not be quantified here.

For 1), rate density, the number of actual events per unit time and sky area is Poisson-distributed.
For 2),  area, a given signal in the time series is superimposed on system noise. In some cases, a signal with S/N~$<7\sigma$ will be detectable (meet the threshold requirements) due to a power increase from the noise power, and conversely a signal with S/N~$>7\sigma$ will be undetectable. The actual number of detected events follows the Poisson distribution.

These stochastic uncertainties are independent, leading to an overall number of detections (in the limit where a Poisson-distributed variable follows the Gaussian distribution):
\begin{equation}
N = \langle{N}\rangle \pm 2\sqrt{N}.
\end{equation}
These uncertainties have been included in the results (Table \ref{results_table}). For cases where the Gaussian distribution is a poor approximation to the Poisson distribution, the Poisson-distributed 68\% confidence intervals have been shown.

\section{Discussion and conclusions}
The MWA is expected to detect a high rate of fast transients events. This is due to its large collecting area and large field-of-view, which, for high S/N events, is sensitive to a large fraction of the sky. In fully coherent mode and under a favourable scattering scenario, tens of events per day may be detectable within the beam. Under conservative scattering conditions, and with computationally feasible modes (incoherent, fast imaging), a few events per week are expected to be detectable. We note again that the evidence suggests that these events match more closely to the zero scattering scenario. These rates are dependent on the underlying assumption that FRBs are standard candles at extragalactic distances, and the host galaxy and our Galaxy play minimal roles in the signal dispersion. The latter assumptions are valid off the plane of the Galaxy, where the electron column is low.

Despite the poorer temporal resolution of the imaging mode, the dispersive smearing of the signal degrades the intrinsic pulse width substantially, degrading the coherent and incoherent modes and making the imaging mode competitive.

Our results are broadly consistent with those presented in \citet{hassall13}. They used a lower intrinsic luminosity for the source population, a simple model for the MWA beam (which did not consider the full beam structure), and considered sub-optimal temporal resolution for the MWA in fast imaging and beamformed modes. Their results may therefore be viewed as lower limits compared with those presented here, for very conservative instrument parameters.

In addition to the detection modes considered here, \citet{law12} \citep[see also][]{kulkarni89} presented the bispectrum technique as an option for FRB detection with interferometric arrays. The bispectrum operates on products of visibilities across three baselines, and yields good performance when the S/N per baseline exceeds unity. For the MWA, this condition will rarely be met, due to the large noise on an individual baseline at high temporal and spectral resolution (MWA SEFD at $\nu=$~200~MHz is $\sim$20~kJy).

Failure of the MWA to detect many events in its first months of fast transients observing will place strong constraints on the assumptions of the source population luminosity function, sky density, and intergalactic electron content. Success of the MWA to detect a large number of events, as predicted in this work, will provide an excellent statistical sample of the FRB population, and allow source localization.  Localization that allows identification of host galaxies with the possibility of determining redshifts would provide an enormous step forward in our understanding of FRBs.

\acknowledgments{The authors thank Jean-Pierre Macquart, Matthew Bailes and the anonymous referee for useful discussions and suggestions. The Centre for All-sky Astrophysics is an Australian Research Council Centre of Excellence, funded by grant CE110001020. The International Centre for Radio Astronomy Research is a Joint Venture between Curtin University and the University of Western Australia, funded by the State Government of Western Australia and the Joint Venture partners. SJT is a Western Australian Premiers Research Fellow.}


\end{document}